\documentclass[12pt]{article}
\usepackage{amssymb,amsmath,amsthm,amsfonts,amscd}
\newcommand\be{\begin{equation}}
\newcommand\ee{\end{equation}}
\newcommand\bea{\begin{eqnarray}}
\newcommand\eea{\end{eqnarray}}

\newcommand{\fatalpha}{{\bf \alpha \kern -0.44em \alpha}}
\newcommand{\fatsigma}{{\bf \sigma \kern -0.54em \sigma}}
\newcommand{\tpchi}{{\bf \chi \kern -0.35em \chi}}
\newcommand{\llambda}{{\bf \lambda \kern -0.45em \lambda}}

\renewcommand{\theequation}{\arabic{equation}}
\renewcommand{\theequation}{\thesection.\arabic{equation}}
\bibliography{plain}
\title{\bf Maximal Entanglement of Two-qubit  States Constructed by Linearly Independent Coherent States} 
 \author{ G. Najarbashi $^{a}$
 \thanks{E-mail: najarbashi@uma.ac.ir},
Y. Maleki $^{a}$
\thanks{E-mail: ymaleki@uma.ac.ir},
 \\ $^{a}${\small Department of Physics, Mohaghegh Ardabili University,
Ardabil 56199-11367, Iran.}}

\begin{document}
\maketitle

\newpage 
\begin{abstract}
In this paper, we find the necessary and sufficient condition for
the maximal entanglement of the  state, $
|\psi\rangle=\mu|\alpha\rangle|\beta\rangle+\lambda|\alpha\rangle|\delta\rangle+
\rho|\gamma\rangle|\beta\rangle+\nu|\gamma\rangle|\delta\rangle,$
constructed by linearly independent coherent states with \emph{real
parameters} when
$\langle\alpha|\gamma\rangle=\langle\beta|\delta\rangle$. This is a
further generalization of the  classified nonorthogonal states
discussed
 in Ref.  Physics Letters A
{\bf{291}}, 73-76 (2001).


{\bf PACs Index: 03.65.Ud }
\end{abstract}
\pagebreak

\vspace{7cm}

\section{Introduction}
Quantum entanglement is one of the most profound features of quantum
mechanics and has been considered to be a valuable physical resource
in the rapidly developing field of quantum information science. In
fact a fundamental difference between quantum and classical physics
is the possible existence of quantum entanglement between distinct
systems. Therefore, manipulation of entangled states is very
important and a challenging problem. By definition, a pure quantum
state of two or more subsystems is said to be entangled if it is not
a product of the states of each component \cite{Nielsen1,petz1}.
Among various entangled states, entanglement of coherent states has
attracted special attention, and due to its applications in quantum
optics \cite{Glauber}, quantum computation and  information,  a lot
of work has been devoted to this problem in the last two decades
\cite{Enk1,Enk2,Fujii1,Wang1,Wang2,Wang3,Wang4,Wang5}.
\par
The required conditions for the maximal  entanglement in superposed bosonic
coherent states of the form
\begin {equation}\label{wang}
|\varphi\rangle=\mu|\alpha\rangle|\beta\rangle+\nu|\gamma\rangle|\delta\rangle,
\end {equation}
have been studied, and the maximally entangled coherent states have
been classified in Ref. \cite{Wang1}. This investigation has been
done using the concurrence measure \cite{Wootters1}. We note that in
the framework of two-qubit states, constructed by linearly
independent coherent states, the above state is a special case of
the state
\begin {equation}\label{gen}
|\psi\rangle=\mu|\alpha\rangle|\beta\rangle+\lambda|\alpha\rangle|\delta\rangle+
\rho|\gamma\rangle|\beta\rangle+\nu|\gamma\rangle|\delta\rangle.
\end {equation}
Note that the above state can not be deduced from the state
(\ref{wang}) by means of local unitary operations $SU(2)\otimes SU(2)$, and consequently
the state (\ref{wang}) is not the Schmidt form of 
(\ref{gen}). This is due to the fact that the coefficients $\mu$ and $\nu$ are generally complex numbers  while in the Schmidt form they must be  real positive numbers. On the other hand, the basis
$|\alpha\rangle$ and $|\gamma\rangle$ (also $|\beta\rangle$ and
$|\delta\rangle$) are not orthonormal, while in Schmidt form they must
be orthonormal. Here we apply the maximal  condition of the
concurrence measure on the previous state with \emph{real
parameters}, and show that in the case $
\langle\alpha|\gamma\rangle=\langle\beta|\delta\rangle$, there would
be only two disjoint classes of maximal entangled coherent states
\\
(\textbf{a}) $\nu=1$ and $\lambda+\rho=-2\langle\alpha|\gamma\rangle$,
\\
(\textbf{b}) $\lambda=\rho$ and $\nu+1=-2\lambda\langle\alpha|\gamma\rangle$.
\section{Entanglement}
A bosonic coherent state is defined as the eigen-state of the
annihilation operator as below
\begin {equation}\label{coherent1}
a|\alpha\rangle=\alpha|\alpha\rangle,
\end {equation}
where $\alpha$ is a complex number, and $a$ is the annihilation
operator for the bosonic harmonic oscillator. Thus, considering the
definition of the coherent state in the number states space, it can
be written as follows
\begin {equation}\label{coherent2}
|\alpha\rangle=e^{\frac{-|\alpha|^2}{2}}
\sum_{n=0}^\infty\frac{\alpha^n}{\sqrt{n!}}|n\rangle=D(\alpha)|0\rangle,
\end {equation}
where $D(\alpha)$ is the displacement operator, and we have
\begin {equation}
D(\alpha):=\exp(a^\dag\alpha-\alpha^*a).
\end {equation}
The overlap of two coherent states is
\begin {equation}\label{overlap}
\langle\alpha|\beta\rangle=e^{-\frac{1}{2}(|\alpha|^2+|\beta|^2-2\alpha^*\beta)}.
\end {equation}
One of the useful measures which can be utilized to quantify the
amount of entanglement in two-qubit states is concurrence, which is
defined as \cite{Wootters1}
\begin {equation}\label{concurrence}
C=|\langle\zeta|\sigma_y\otimes\sigma_y|\zeta^*\rangle|,
\end {equation}
where $\sigma_y$ is the $y$ component of the usual Pauli spin
matrices. Now let us consider the general form of the two-qubit
coherent state i.e.,
\begin {equation}\label{general eq}
|\psi\rangle=\mu|\alpha\rangle|\beta\rangle+\lambda|\alpha\rangle|\delta\rangle+
\rho|\gamma\rangle|\beta\rangle+\nu|\gamma\rangle|\delta\rangle,
\end {equation}
where $|\alpha\rangle$ and $|\gamma\rangle$ are linearly independent
normalized states, which span the space of the system 1, and
$|\beta\rangle$ and $|\delta\rangle$ are linearly independent
normalized states, which span the space of the system 2. The
normalization of the state (\ref{general eq}) is
$$
N^2=\langle\psi|\psi\rangle=|\mu|^2+\mu^*\lambda
\langle\beta|\delta\rangle+\mu^*\rho
\langle\alpha|\gamma\rangle+\mu^*\nu\langle\alpha|\gamma\rangle\langle\beta|\delta\rangle+
\lambda^*\mu\langle\beta|\delta\rangle^*+|\lambda|^2+
$$
$$
\lambda^*\rho\langle\alpha|\gamma\rangle\langle\beta|\delta\rangle^*+
\lambda^*\nu\langle\alpha|\gamma\rangle+\rho^*\mu\langle\alpha|\gamma\rangle^*+
\rho^*\lambda\langle\alpha|\gamma\rangle\langle\beta|\delta\rangle+|\rho|^2+
$$
\begin {equation}
\rho^*\nu\langle\beta|\delta\rangle+\nu^*\mu\langle\alpha|\gamma\rangle^*\langle\beta|\delta\rangle^*+
\nu^*\lambda\langle\alpha|\gamma\rangle^*+\nu^*\rho\langle\beta|\delta\rangle^*+|\nu|^2.
\end {equation}
In the case $\rho=\lambda=0$, $|\psi\rangle$ reduces to the state
discussed in \cite{Wang1}. Here we are going to develop the
discussion to some general form in \emph{real field}. We assume,
without loss of generality, that $\mu=1$ and get
\begin {equation}\label{general eq2}
|\psi\rangle=|\alpha\rangle|\beta\rangle+\lambda|\alpha\rangle|\delta\rangle+
\rho|\gamma\rangle|\beta\rangle+\nu|\gamma\rangle|\delta\rangle.
\end {equation}
\par
\textbf{Theorem:} The general form of the two-qubit
 state Eq. (\ref{general eq2}) constructed by linearly independent coherent states  in real field with requirement
$\langle\alpha|\gamma\rangle=\langle\beta|\delta\rangle$, is
maximally entangled if and only if one of the following conditions
holds
\\
(\textbf{a}) $\nu=1$ and
$\lambda+\rho=-2\langle\alpha|\gamma\rangle$,
\\
(\textbf{b}) $\lambda=\rho$ and
$\nu+1=-2\lambda\langle\alpha|\gamma\rangle$.
\par
In order to prove the theorem, we take the following orthonormal
basis
$$
|0\rangle=|\alpha\rangle, \quad
|1\rangle=\frac{|\gamma\rangle-p_1|\alpha\rangle}{N_1} \quad
\text{for system} 1,
$$
\begin {equation}
|0\rangle=|\delta\rangle, \quad
|1\rangle=\frac{|\beta\rangle-p_2|\delta\rangle}{N_2} \quad
\text{for system} 2,
\end {equation}
where
$$
p_1=\langle\alpha|\gamma\rangle, \quad  N_1=\sqrt{1-|p_1^2|},
$$
\begin {equation}
p_2=\langle\delta|\beta\rangle, \quad  N_1=\sqrt{1-|p_2^2|}.
\end {equation}
Taking these new bases we can rewrite $|\psi\rangle$ as
\begin {equation}\label{general base}
|\psi\rangle=\frac{(p_2+\lambda+\rho p_1p_2+\nu
p_1)|00\rangle+N_2(1+\rho p_1)|01\rangle+N_1(\nu+\rho
p_2)|10\rangle+\rho N_1N_2|11\rangle}{N}.
\end {equation}
For a general two-qubit pure state
\begin {equation}
|\varphi\rangle=a|00\rangle+b|01\rangle+c|10\rangle+d|11\rangle,
\end {equation}
the concurrence is given by
\begin {equation}\label{concurrecce}
C=2|ad-bc|.
\end {equation}
Using Eqs. (\ref{general base}) and (\ref{concurrecce}), the
concurrence of the state (\ref{general eq2}) becomes
\begin {equation}\label{concurrecce2}
C=\frac{2|\nu-\lambda\rho|(\sqrt{1-|p_1^2|})(\sqrt{1-|p_2^2|})}{N^2}.
\end {equation}
We just survey the case
$\langle\alpha|\gamma\rangle=\langle\beta|\delta\rangle $ in detail.
Then the Eq. (\ref{overlap}) for $\alpha$, $\gamma\in \mathbb{R}$
becomes
\begin {equation}
0<\langle\alpha|\gamma\rangle=e^{\frac{-1}{2}(\alpha-\gamma)^2}<1.
\end {equation}
By solving Eq. (\ref{concurrecce2}) for $C=1$ which is equivalent to
maximal entanglement of the state (\ref{general eq2}), we get
\begin {equation}\label{eq1}
2|\nu-\lambda\rho|(1-x^2)=(1+\lambda^2+\rho^2+\nu^2)+2(\lambda+\rho+\nu\lambda+\rho\nu)x+2(\nu+\lambda\rho)x^2.
\end {equation}
We divide this equation into two parts as follows
\par
\textbf{Case 1:} Let $\nu>\lambda\rho$,
then we have
\begin {equation}
(1-\nu)^2+(\lambda+\rho)^2+2(\lambda+\rho)(1+\nu)x+4\nu x^2=0,
\end {equation}
this equation is a second order polynomial with respect to $x$ and
has solutions if and only if $\nu=1$ and $\lambda+\rho=-2x$ (see appendix).
\\
For example, the following states are maximally entangled states and
belong to this class
\begin {equation}
|\psi\rangle_{max}=\frac{1}{\sqrt{2(1-e^{-(\alpha-\gamma)^2})}}\left(|\alpha\rangle|\beta\rangle-e^{\frac{-1}{2}(\alpha-\gamma)^2}
|\alpha\rangle|\delta\rangle-e^{\frac{-1}{2}(\alpha-\gamma)^2}
|\gamma\rangle|\beta\rangle+|\gamma\rangle|\delta\rangle\right),
\end {equation}
\begin {equation}
|\psi\rangle_{max}=\frac{1}{\sqrt{2(1-e^{-(\alpha-\gamma)^2})}}\left(|\alpha\rangle|\beta\rangle-2e^{\frac{-1}{2}(\alpha-\gamma)^2}
|\alpha\rangle|\delta\rangle+|\gamma\rangle|\delta\rangle\right),
\end {equation}
\begin {equation}
|\psi\rangle_{max}=\frac{1}{\sqrt{2(1+2e^{-(\alpha-\gamma)^2})-3e^{-2(\alpha-\gamma)^2}}}\left(|\alpha\rangle|\beta\rangle+e^{\frac{-1}{2}(\alpha-\gamma)^2}
|\alpha\rangle|\delta\rangle-3e^{\frac{-1}{2}(\alpha-\gamma)^2}
|\gamma\rangle|\beta\rangle+|\gamma\rangle|\delta\rangle\right),
\end {equation}
Taking $\beta=\alpha$ and $\delta=\gamma$ in the above states, we
get
\begin {equation}
|\psi\rangle_{max}=\frac{1}{\sqrt{2(1-e^{-(\alpha-\gamma)^2})}}\left(|\alpha\rangle|\alpha\rangle-e^{\frac{-1}{2}(\alpha-\gamma)^2}
|\alpha\rangle|\gamma\rangle-e^{\frac{-1}{2}(\alpha-\gamma)^2}
|\gamma\rangle|\alpha\rangle+|\gamma\rangle|\gamma\rangle\right),
\end {equation}
\begin {equation}
|\psi\rangle_{max}=\frac{1}{\sqrt{2(1+2e^{-(\alpha-\gamma)^2})-3e^{-2(\alpha-\gamma)^2}}}\left(|\alpha\rangle|\alpha\rangle+e^{\frac{-1}{2}(\alpha-\gamma)^2}
|\alpha\rangle|\gamma\rangle-3e^{\frac{-1}{2}(\alpha-\gamma)^2}
|\gamma\rangle|\alpha\rangle+|\gamma\rangle|\gamma\rangle\right),
\end {equation}
\begin {equation}
|\psi\rangle_{max}=\frac{1}{\sqrt{2(1-e^{-(\alpha-\gamma)^2})}}\left(|\alpha\rangle|\alpha\rangle-2e^{\frac{-1}{2}(\alpha-\gamma)^2}
|\alpha\rangle|\gamma\rangle+|\gamma\rangle|\gamma\rangle\right).
\end {equation}
\textbf{Case 2:} Let $\nu<\lambda\rho$,
then we have
\begin {equation}\label{eq2}
(1+\nu)^2+(\lambda-\rho)^2+2(\lambda+\rho)(1+\nu)x+4\lambda\rho
x^2=0.
\end {equation}
The above equation has solutions if and only if $\lambda=\rho$ and
$\nu+1=-2\lambda x$. For example the following states are maximally
entangled states and belong to this class
\begin {equation}
\hspace{-2cm}|\psi\rangle_{max}=\frac{1}{\sqrt{2(1-e^{-(\alpha-\gamma)^2})}}\left(|\alpha\rangle|\beta\rangle-|\gamma\rangle|\delta\rangle\right),
\end {equation}
\begin {equation}
\hspace{-2cm}|\psi\rangle_{max}=\frac{e^{\frac{-1}{2}(\alpha-\gamma)^2}}{\sqrt{2(1-e^{-(\alpha-\gamma)^2})}}\left(|\alpha\rangle|\beta\rangle-
e^{\frac{1}{2}(\alpha-\gamma)^2}|\alpha\rangle|\delta\rangle-e^{\frac{1}{2}(\alpha-\gamma)^2}
|\gamma\rangle|\beta\rangle+|\gamma\rangle|\delta\rangle\right),
\end {equation}
\begin {equation}
\hspace{-1.9cm}|\psi\rangle_{max}=\frac{e^{\frac{-1}{2}(\alpha-\gamma)^2}}{\sqrt{2(4-7e^{-(\alpha-\gamma)^2}+3e^{-2(\alpha-\gamma)^2})}}\left(|\alpha\rangle|\beta\rangle-2
e^{\frac{1}{2}(\alpha-\gamma)^2}|\alpha\rangle|\delta\rangle-2e^{\frac{1}{2}(\alpha-\gamma)^2}
|\gamma\rangle|\beta\rangle+3|\gamma\rangle|\delta\rangle\right),
\end {equation}
\begin {equation}
\hspace{-2cm}|\psi\rangle_{max}=\frac{e^{\frac{-1}{2}(\alpha-\gamma)^2}}{\sqrt{2(1+2e^{-(\alpha-\gamma)^2}-3e^{-2(\alpha-\gamma)^2})}}\left(|\alpha\rangle|\beta\rangle+
e^{\frac{1}{2}(\alpha-\gamma)^2}|\alpha\rangle|\delta\rangle+e^{\frac{1}{2}(\alpha-\gamma)^2}
|\gamma\rangle|\beta\rangle-3|\gamma\rangle|\delta\rangle\right),
\end {equation}
\begin {equation}
\hspace{-2.5cm}|\psi\rangle_{max}=\frac{\sqrt{2}e^{\frac{-1}{2}(\alpha-\gamma)^2}}{\sqrt{(1-e^{-(\alpha-\gamma)^2})}}\left(|\alpha\rangle|\beta\rangle-\frac{1}{2}
e^{\frac{1}{2}(\alpha-\gamma)^2}|\alpha\rangle|\delta\rangle-\frac{1}{2}e^{\frac{1}{2}(\alpha-\gamma)^2}
|\gamma\rangle|\beta\rangle\right).
\end {equation}
The first state is the antisymmetric maximally entangled state
obtained in \cite{Wang5}. The proof of the inverse of the theorem,
is straightforward and can be verified directly by putting the above
conditions in the concurrence measure.
\par
Now let us consider some special cases of the theorem. For example
when $\langle\alpha|\gamma\rangle\rightarrow0$ (satisfied when
$|\alpha-\gamma|\rightarrow \infty$) we get from the condition
(\textbf{a}) that the following  state is maximal
\begin {equation}
\frac{1}{\sqrt{2(1+\lambda^2)}}(\lambda|00\rangle+|01\rangle+|10\rangle-\lambda|11\rangle),
\end {equation}
and  taking
$\lambda=0$, the state reduces to one of the
 Bell states, i.e.,
\begin {equation}
\frac{1}{\sqrt{2}}(|01\rangle+|10\rangle).
\end {equation}
However taking the condition (\textbf{b}) we have
\begin {equation}
\frac{1}{\sqrt{2(1+\lambda^2)}}(\lambda|00\rangle+|01\rangle-|10\rangle+\lambda|11\rangle),
\end {equation}
which taking $\lambda=0$,
reduces to another  Bell state, i.e.,
\begin {equation}
\frac{1}{\sqrt{2}}(|01\rangle-|10\rangle).
\end {equation}
\\
\textbf{Remark:} The state (\ref{general eq}) with $\mu=1$ is
separable (C = 0) if and only if  $\nu=\lambda\rho.$

We note that the separability  conditions hold  in  complex field
too. For example the following states, up to normalization factors,
are separable
\begin {equation}
\hspace{-.9cm}|\psi\rangle_{sep}=|\alpha\rangle|\beta\rangle+|\alpha\rangle|\delta\rangle+
|\gamma\rangle|\beta\rangle+|\gamma\rangle|\delta\rangle,
\end {equation}
\begin {equation}
|\psi\rangle_{sep}=|\alpha\rangle|\beta\rangle+\lambda|\alpha\rangle|\delta\rangle+
\rho|\gamma\rangle|\beta\rangle+\lambda\rho|\gamma\rangle|\delta\rangle,
\end {equation}
\begin {equation}
\hspace{-.9cm}|\psi\rangle_{sep}=|\alpha\rangle|\beta\rangle-|\alpha\rangle|\delta\rangle+
|\gamma\rangle|\beta\rangle-|\gamma\rangle|\delta\rangle,
\end {equation}
\begin {equation}
|\psi\rangle_{sep}=|\alpha\rangle|\beta\rangle+|\alpha\rangle|\delta\rangle+
\rho|\gamma\rangle|\beta\rangle+\rho|\gamma\rangle|\delta\rangle.
\end {equation}
\par

In summery, we investigated the  entanglement of two-qubit coherent
states $
|\psi\rangle=\mu|\alpha\rangle|\beta\rangle+\lambda|\alpha\rangle|\delta\rangle+
\rho|\gamma\rangle|\beta\rangle+\nu|\gamma\rangle|\delta\rangle,$
with real parameters, assuming
$\langle\alpha|\gamma\rangle=\langle\beta|\delta\rangle$. This
problem has been solved for the special case
$|\varphi\rangle=\mu|\alpha\rangle|\beta\rangle+\nu|\gamma\rangle|\delta\rangle$
in complex field in Ref. \cite{Wang1}. However, finding all maximal
entangled  regions of the Eq. (\ref{general eq}) in complex field,
is challenging and open for debate.

\vspace{1cm} \setcounter{section}{0}
 \setcounter{equation}{0}
 \renewcommand{\theequation}{\arabic{equation}}
\newpage
{\Large{Appendix :}}\\
In this appendix, we show that the  equation
\begin {equation}
(1-\nu)^2+(\lambda+\rho)^2+2(\lambda+\rho)(1+\nu)x+4\nu x^2=0,
\end {equation}
with constraint  $0<x<1$ has solution iff both $\nu=1$ and $\lambda+\rho=-2\langle\alpha|\gamma\rangle$, hold.
We note that, if all the coefficients of the left hand side vanish, the
equality  holds identically and it would become independent of $x$, but
here we can not take all the coefficients to be zero.
Now we suppose that $\nu=0$, then
\begin {equation}
x=\frac{1+(\lambda+\rho)^2}{-2(\lambda+\rho)},
\end {equation}
applying $0<x<1$, yields $(1+\lambda+\rho)^2<0$ which is impossible. Thus  $\nu\neq0$. The above second order polynomial has solutions
\begin {equation}
x_\pm=\frac{-(\lambda+\rho)(1+\nu)\pm\sqrt{(1-\nu)^2((\lambda+\rho)^2-4\nu)}}{4\nu},
\end {equation}
Since $x_\pm$ are real parameters, then
$(1-\nu)^2((\lambda+\rho)^2-4\nu)\geq0$. We discuss equality and inequality cases separately.
\\
\textbf{I:}
\\
 Let $(1-\nu)^2((\lambda+\rho)^2-4\nu)=0$, then
 $(\lambda+\rho)^2=4\nu$, or $\nu=1$. If $(\lambda+\rho)^2=4\nu$
 then
$$
0<x_\pm=\frac{-(\lambda+\rho)(1+\nu)}{4\nu}<1\quad
\Longrightarrow\quad (\lambda+\rho)<0,
$$
which implies that $(1-\nu)^2<0$ and this is impossible in the real
field. Thus we turn  to the other possibility and
assume that
 $$
 \nu=1 \quad \Longrightarrow \quad 0<x_\pm=\frac{-(\lambda+\rho)}{2}<1\quad
\Longrightarrow\quad , \lambda+\rho=-2\langle\alpha|\gamma\rangle,
$$
which is the result (\textbf{a}) of the main theorem.
\\
\textbf{II:}
\\
 Now let us survey the case
$(1-\nu)^2((\lambda+\rho)^2-4\nu)>0$ or equivalently
\begin {equation}\label{frac}
(\lambda+\rho)^2>4\nu \quad \textrm{and}\quad \nu\neq1
\end {equation}
 We discuss two solutions  $x_\pm$ for the case  $\nu>0$. The same discussion holds if $\nu<0$.
 First let us concentrate on $x_-$,
thus
\begin {equation}\label{ap1}
x_->0 \quad \Longrightarrow\quad (\lambda+\rho)<-2\sqrt{\nu},
\end {equation}
\begin {equation}\label{}
x_-<1 \quad\Longrightarrow\quad
-((\lambda+\rho)(1+\nu)+4\nu)<\sqrt{(1-\nu)^2((\lambda+\rho)^2-4\nu)}.
\end {equation}
If $-((\lambda+\rho)(1+\nu)+4\nu,$ belongs to the following bounds
\begin {equation}\label{}
-\sqrt{(1-\nu)^2((\lambda+\rho)^2-4\nu)}<-((\lambda+\rho)
(1+\nu)+4\nu)<\sqrt{(1-\nu)^2(\lambda+\rho)^2-4\nu)},
\end {equation}
then one easily gets $(1+\lambda+\rho)^2<0$ which is a contradiction. Therefore  we
must restrict ourselves  to the  following bound
\begin {equation}\label{}
-((\lambda+\rho)
(1+\nu)+4\nu)\leq-\sqrt{(1-\nu)^2((\lambda+\rho)^2-4\nu)}.
\end {equation}
This equation implies that
\begin {equation}\label{ap2}
(\lambda+\rho)>\frac{-4\nu}{1+\nu} ,
\end {equation}
The two Eqs. (\ref{ap1}) and
(\ref{ap2}) yield $(\nu-1)^2<0$ and consequently, we
have no solution for $x_-$ in this case.
\par
Now we turn our attention to $x_+$, thus
\begin {equation}\label{ineq1}
0<x_+=\frac{-(\lambda+\rho)(1+\nu)+\sqrt{(1-\nu)^2((\lambda+\rho)^2-4\nu)}}{4\nu}<1,
\end {equation}
The upper bound yields
\begin {equation}
x_+<1 \quad\Longrightarrow \quad(\lambda+\rho)>\frac{-4\nu}{1+\nu}
\end {equation}
which  together with Eq. (\ref{frac}) reads
$$
(\lambda+\rho)>2\sqrt{\nu}.
$$
and the lower bound of Eq. (\ref{ineq1}) yields
\begin {equation}
(\lambda+\rho)(1+\nu)<\sqrt{(1-\nu)^2((\lambda+\rho)^2-4\nu)},
\end {equation}
which leads to
\begin {equation}
(\lambda+\rho)^2+(1-\nu)^2<0.
\end {equation}
which is evidently a contradiction.
\par
One can pursue the same line as above and consider the solutions of Eq. (\ref{eq2})
in order to obtain the result (\textbf{b}) of the main theorem.

\end{document}